\RequirePackage{fix-cm}
\documentclass[twocolumn,epjc3]{svjour3}
\smartqed
\RequirePackage{graphicx}
\RequirePackage{mathptmx}
\RequirePackage{amsmath}
\RequirePackage{amsfonts}
\RequirePackage{amssymb}
\RequirePackage{latexsym}
\RequirePackage{widetext}

\newcommand{\be}{\begin{equation}}
\newcommand{\ee}{\end{equation}}
\newcommand{\bea}{\begin{eqnarray}}
\newcommand{\eea}{\end{eqnarray}}

\journalname{Eur. Phys. J. C}
\begin{document}

\title{Violation of causality in $f(T)$ gravity 
}

\author{G. Otalora\thanksref{e1,addr1}
        \and
        M. J. Rebou\c{c}as\thanksref{e2,addr2}}

\thankstext{e1}{e-mail: giovanni.otalora@pucv.cl}
\thankstext{e2}{e-mail: reboucas@cbpf.br}

\institute{Instituto de F\'{\i}sica, Pontificia Universidad Cat\'olica de
Valpara\'{\i}so, Casilla 4950, Valpara\'{\i}so, Chile \label{addr1}
           \and
Centro Brasileiro de Pesquisas F\'{\i}sicas,
Rua Dr.\ Xavier Sigaud 150,
22290-180 Rio de Janeiro -- RJ, Brazil\label{addr2}
}

\date{Received: date / Accepted: date}

\maketitle

\sloppy

\begin{abstract}
In its standard formulation, the $f(T)$ field equations are not invariant
under local Lorentz transformations, and thus the theory does not inherit the
causal structure of special relativity.
Actually, even locally violation of causality can occur in this formulation
of $f(T)$ gravity.
A locally Lorentz covariant $f(T)$ gravity theory  has been devised recently,
and this local causality problem seems to have been overcome. The nonlocal question,
however, is left open.
If gravitation is to be described by this covariant $f(T)$ gravity theory
there are a number of issues that ought to be examined in its context,
including the question as to whether its field equations allow homogeneous
G\"odel-type solutions, which necessarily leads to violation of causality
on nonlocal scale.
Here, to look into the potentialities and difficulties of the covariant
$f(T)$ theories, we examine whether they admit G\"odel-type solutions.
We take a combination of a perfect fluid with electromagnetic plus a scalar
field as source, and determine a general G\"odel-type solution, which
contains special solutions in which the essential parameter
of G\"odel-type geometries, $m^2$, defines any class of homogeneous
G\"odel-type geometries.
We show that solutions of  the trigonometric and linear classes ($m^2 < 0$
and $m=0$) are permitted only for the combined matter sources with an
electromagnetic field matter component.
We extended to the context of covariant $f(T)$  gravity a theorem, which
ensures that any perfect-fluid homogeneous G\"odel-type solution
defines the same set of G\"odel tetrads $h_A^{~\mu}$ up to a Lorentz transformation.
We also showed  that the single massless scalar field generates G\"odel-type
solution with no closed timelike curves.
Even though the covariant $f(T)$ gravity restores Lorentz covariance of
the field equations and the local validity of the causality principle,
the bare existence of the G\"odel-type solutions makes apparent that the
covariant formulation of $f(T)$ gravity does not preclude non-local violation
of causality in the form of closed timelike curves.
\end{abstract}


\section{Introduction}\label{Introduction}

The frameworks proposed to account for the observed late-time accelerated expansion
of the Universe can be roughly grouped into two broad families.
In the first, the unknown form of matter sources, the so-called dark energy, is invoked
and the underlying gravity theory, general relativity (GR), is kept unmodified.
In this framework, the simplest way to describe the accelerated expansion of the
Universe is by introducing a cosmological constant, $\Lambda$, into the general
relativity field equations.
This approach is entirely consistent with the available observational data, but
it faces difficulties related to  the order of magnitude of the cosmological
constant and its microphysical origin.
In the second family, modifications of Einstein's field equations are taken
as an alternative for describing the late-time accelerated expansion.
Examples in this class include generalized gravity theories based upon
modifications of the Einstein-Hilbert action by taking nonlinear functions $f(R)$
of the Ricci scalar $R$  or other curvature invariants (for reviews see, e.g.,
Refs.~\cite{fr-reviews1,fr-reviews2,fr-reviews3,fr-reviews4,fr-reviews5,%
fr-reviews6,fr-reviews7}).

Another family of modified gravity theories that has been examined
as an alternative way of describing the late-time acceleration of the
Universe~\cite{Bengochea09,Linder10,Myrzakulov11,Cai:2015emx} is known
as $f(T)$ gravity.
In close analogy with the $f(R)$,  the $f(T)$ gravity theory
was suggested by extending the Lagrangian of teleparallel gravity to a
function $f(T)$ of a torsion scalar $T$~\cite{Bengochea09,Linder10}.
The Lagrangian density in these so-called teleparallel gravity (TG) theories
contains a function $f(T)$ of the torsion scalar $T$ in a formal analogy
with $f(R)$ gravity.%
\footnote{
The very beginning of this subject goes back to Einstein's attempt to unify gravity
and electromagnetism in 1928 through the introduction of vierbein (tetrad) field
together with the notion of distant parallelism or teleparallelism%
~\cite{Einstein,TranslationEinstein}.}
In the teleparallel gravity theories the dynamical variables are the tetrad fields
$\mathbf{h}^{}_{A}(x^\mu)$ instead of the metric $g_{\mu \nu}$, and rather than
the usual torsionless Levi-Civita connection of GR, a Weitzenb\"{o}ck connection,
which has torsion but no curvature, is employed to define the covariant derivative.
In this way, the gravitational field in teleparallel gravity is then described
in terms of the torsion instead of the curvature%
~\cite{Early-papers1,Early-papers3,Early-papers4,Early-papers5,%
Early-papers6,Aldrovandi-Pereira-book,AndradeGuillenPereira-00}.   

In comparison with $f(R)$ in the metric formalism, whose field equations are
of fourth-order, the $f(T)$ gravity theories have the advantage that their
dynamics are given by second-order differential equations.
This important characteristic, along with the fact that $f(T)$ theories can
be used to explain the observed accelerating expansion, has given birth to a
fair number of articles on these gravity theories, in which several features
of $f(T)$ gravity have been examined, including observational solar system
constraints~\cite{Iorio-Saridakis-2012,Iorio-2015,Farrugia-2016}, cosmological
constraints~\cite{Bengochea-2011,Wei-Ma-Qi-2011,Capozziello-Luongo-Saridakis-2015,%
Oikonomou-Saridakis-2016,Nunes-Pan-Saridakis-2016},
cosmological perturbations~\cite{Dent-Duta-Saridakis-2011,Zheng-Huang-2011,Izumi:2012qj,%
Li:2011wu,Basilakos:2016xob}, spherically symmetric solutions~\cite{Wang-2011,%
Atazadeh:2012am,Ruggiero:2015oka}, the existence of relativistic stars~\cite{Stars-in-f(T)},
cosmographic constraints~\cite{Cosmography-2011} and energy conditions
bounds~\cite{Liu-Reboucas-2012}.%
\footnote{For further references on several aspects of $f(T)$ gravity  we refer the readers
to the review article~\cite{Cai:2015emx} and references therein. For an incomplete list of
more recent references is
~\cite{FT1,FT3,FT4,FT5,FT9,FT10,FT11,FT12,FT14,FT15,FT16,FT17,FT18,FT19,%
FT21,FT24,FT26,FT28,FT36,FT37,FT38}.}

Despite this noticeable interest in the new gravity theory, it has been pointed out
that the field equations of $f(T)$ theory are not invariant under local Lorentz
transformations~\cite{Li-Sotiriou-Barrow-2011,Miao-Li-Miao-2011} unless the $f(T)$
gravity is teleparallel equivalent of general relativity (TEGR), in which
$f(T) = \lambda\, T + \Lambda$, where $\lambda$ and $\Lambda$ are constants.
This means that apart from the TEGR, these theories are, in general, sensitive to the
choice of the tetrads --- Lorentz transformation of tetrad basis changes the $f(T)$
field equations~\cite{Sotiriou-Li-Barrow2011b,Zheng2011,Tamanini-Bohmer-2012}.
This problem appears to have been initially overlooked in most of recent literature
on $f(T)$ gravity, but has recently been considered in Ref.~\cite{Krssak:2015oua},
where it has been found that lack of Lorentz invariance arises as a consequence of
a particular choice of the spin connection borrowed from TEGR~\cite{Krssak:2015oua}.
Relaxing this original choice of the spin connection, a locally Lorentz
covariant $f(T)$ gravity theory has then been devised~\cite{Krssak:2015oua}.

In the special relativity  chronology and causality are so essential
ingredients that they are simply built into the theory from the very outset
--- chronology is preserved and causality is respected in the theory.
General relativity (GR) inherits locally a chronology protection, which ensures
the local validation of the causality principle, from the bare
fact that the space-times of general relativity are locally Minkowskian.
On nonlocal scale, however, important differences may emerge since
Einstein's field equations do not provide in general nonlocal constraints
on the underlying space-time manifold. 
The model found by G\"odel~\cite{Godel49} is a well-known solution to 
Einstein's equations that makes it clear that GR admits solutions with closed
timelike world lines, despite its local invariance under Lorentz transformations
that ensures the validity of the causality principle locally.%
\footnote{G\"odel's solution has a recognizable importance and has motivated
a fair number of investigations on rotating G\"odel-type models in the context
of general relativity (see, e.g.               Refs.~\cite{GT_in_GR1,GT_in_GR2,GT_in_GR3,GT_in_GR5,GT_in_GR6%
,GT_in_GR7,GT_in_GR10,GT_in_GR11}) and in the framework of
other gravity theories (see, for example, Refs.~\cite{GT_Other-Th4,GT_Other-Th5,%
GT_Other-Th6,GT_Other-Th7,GT_Other-Th8,GT_Other-Th9,GT_Other-Th10,GT_Other-Th11,%
GT_Other-Th12,GT_Other-Th13,GT_Other-Th15,GT_Other-Th16,%
GT_Other-Th18,GT_Other-Th19,GT_Other-Th20}).
}

Even though different choices of tetrads related by a Lorentz transformation
leave the metric invariant, suggesting at first sight a well-defined local causal
structure, in the usual or standard formulation of $f(T)$ gravity every different
Lorentz tetrads  give rise to different field equations, and therefore represents
a different theory.
In this way, for a given (fixed) frame of tetrads, which determines a given (fixed)
$f(T)$ theory, there is no Lorentz transformation freedom. This means that unlike
general relativity, a $f(T)$ gravity theory does not inherit locally a chronology
protection from the special relativity, and one may even have a local violation
of causality.%
\footnote{The local violation causality is also related to ill-defined Cauchy
problem in f(T) gravity. These issues have been discussed in Refs.~\cite{YenChinOng-etal1,YenChinOng-etal2,YenChinOng-etal3}.}
This local causality problem seems to have been overcome in the Lorentz
covariant $f(T)$ gravity theory~\cite{Krssak:2015oua}, since in this new formulation
of the theory the Lorentz transformations do not change neither the metric nor the
field equations.
A question that naturally arises here is whether the covariant formulation of $f(T)$
gravity allows G\"odel-type solutions, which necessarily lead to \textit{nonlocal}
violation of the causality principle in the form of closed timelke curves,
or would remedy this causal pathology by ruling out this type of solutions,
which are permitted in general relativity. Moreover, if gravitation
is to be described by this Lorentz covariant $f(T)$ theory~\cite{Krssak:2015oua}
there are a number of issues that ought to be reexamined in its context,
including the question as to whether these gravity theories allow noncausal
solutions for physically well-behaved matter sources. 

In this paper, we proceed with further investigations on the difficulties, limitations
and potentialities of the Lorentz covariant $f(T)$ theory~\cite{Krssak:2015oua},
by undertaking this question and examining whether this gravity theory admits
homogeneous G\"odel-type solutions for a quite general matter source.
To this end,  we take a combination of a perfect fluid with electromagnetic plus
a scalar fields as a matter source, and determine a general G\"odel-type solution
of the covariant $f(T)$ field equations. This general solution contains several special
solutions, including a perfect-fluid and a single scalar field solutions.
It emerges from our results that this general solution contains special solutions
whose essential parameter, $m$, of G\"odel-type geometries is positive
(hyperbolic family), null (linear family) and negative (trigonometric class).
Solutions in the trigonometric class only exist for combination of  sources
that includes an electromagnetic field as a matter component.
We show that any perfect-fluid
ST-homogeneous G\"odel-type solution of the
covariant $f(T)$ gravity is isometric to G\"odel metric, and therefore exhibits
violation of causality.
\footnote{
This extends to the context of covariant $f(T)$  gravity the Bampi-Zordan
theorem~\cite{BampiZordan78}, which states that every perfect-fluid G\"odel-type
solution to the Einstein field equations is necessarily isometric to the
G\"odel space-time.}
We also show that a single massless scalar field generates the only ST-homogeneous
G\"odel-type solution without violation of causality of G\"odel type.
We underline that the bare existence of these ST-homogeneous G\"odel-type
solutions makes apparent that the Lorentz covariant $f(T)$ gravity does not
remedy causal anomaly in the form of closed timelike curves that
are present in general relativity.

The structure of the paper is as follows. In Section~\ref{CovF(T)}, to define
the notation and make this paper to a certain extent self-contained, we give
a brief account of the Lorentz covariant $f(T)$ gravity.
In Section~\ref{Sec-GodelType} we present the main properties of space-time (ST)
homogenous G\"{o}del-type geometries. This includes the metric, the
ST-homogeneity conditions, the non-isometric ST-homogeneous G\"odel-type classes,
and the existence of closed time-like curves in these space-times.
In Section~\ref{GodelTypeinCovF(T)} we show that the Lorentz covariant  $f(T)$
gravity theories admit ST-homogeneous G\"{o}del-type solutions for several
physically well-motivated matter contents, and therefore despite the local
Lorentz invariance, it houses nonlocal violation of causality.
In Section~\ref{Conclu} we present our main conclusions and final remarks.

\section{Covariant $f(T)$ gravity}\label{CovF(T)}

In this Section we briefly introduce the $f(T)$ gravity theory and its covariant
formulation. For more details we refer the readers to the book~\cite{Aldrovandi-Pereira-book},
where the notation, basic definition and proofs are presented, and to the
article~\cite{Krssak:2015oua}, where the formulation of the covariant version
of $f(T)$ gravity theory is presented. For a pedagogical presentation  of the
basic geometrical setting of these theories see Appendix J of Sean Carroll's
book~\cite{Carroll}.

We begin by recalling that the dynamical variables in $f(T)$ gravity
theories are the tetrad fields, $\mathbf{h}^{}_{A}(x^\mu)$, which is a set of four
($A= 0,\cdots,3$) vector fields that define a local orthonormal Lorentz frame
at every point $x^\mu$ of the spacetime manifold. The vector fields
$\mathbf{h}^{}_{A}(x^\mu)$ are vectors in the tangent space at a arbitrary point
$x^{\mu}$ of the spacetime manifold. The spacetime and the tangent space
metrics are related by
\begin{equation}  \label{metrics}     
 g_{\mu\nu}=h^{A}_{~\mu}\,h^{B}_{~\nu}\,\eta_{AB}^{}\,,
 \end{equation}
where $\eta _{AB}^{}=\text{diag}\,(1,-1,-1,-1)$ is the Minkowski metric of the tangent
space at $x^\mu$. Here and in what follows, we use Greek letters to denote spacetime
coordinate indices, which are raised and lowered, respectively,  with $g_{\mu \nu }$
and $g^{\mu \nu }$ and vary from $0$ to $3$,  whereas Latin upper case letters
denote tetrad indices, which are lowered and raised with the Minkowski tensor
$\eta_{AB}$ and $\eta^{AB}$.
It follows from equation~\eqref{metrics} that the relation between frame components,
$h_A^{~\mu }$,  and coframe components, $h^{A}_{~\mu }$, are given by
\begin{equation}\label{tetradralation}
h_{A}^{~\mu }\,h^{A}_{~\nu}=\delta _{\nu }^{\mu } \qquad \text{and} \qquad
h_{A}^{~\mu}\,h^{B}_{~\mu}=\delta _{A}^{B}\,.
\end{equation}

Under local Lorentz transformations at each point $x$ the tetrad fields
transforms as
\begin{equation}  \label{h_Lorentz-transf} 
\qquad \mathbf{h}^{}_{B} \longrightarrow \mathbf{h'}_{A} =
\Lambda^B{}_{A}\,(x) \,\, \mathbf{h}^{}_{B}
\end{equation}
and leave the metric invariant, i.e.
\begin{equation}
\quad \eta'_{AB} = \Lambda^C{}_{A} \,\, \eta_{CD} \,\, \Lambda^D{}_{B}
= \text{diag}\,(1,-1,-1,-1)\,.
\end{equation}

Considering that at every point of the spacetime besides the general coordinate
transformations, $x^\mu \rightarrow x^{\mu'}$, we have the freedom to perform a
Lorentz transformation one has that a mixed transformation law as, for example
\begin{equation}
M'^{A\mu}_{\;\;\;\;\;\;\;B\nu}=\Lambda^{A}{}_{C}\,\,
              \frac{\partial x^{\mu'}}{\partial x^\lambda}\,\,
              \Lambda^D{}_{B}\,\, \frac{\partial x^\rho}{\partial x^{\nu'}}
               \,\,\,M^{C\lambda}_{\;\;\;\;\;\;D\rho} \,.
\end{equation}
For differentiating geometrical objects as, e.g., $X^A_{\;\;B}$ we replace the ordinary
connection  by the spin connection, denoted by  
$\omega^{A}_{~B \mu}$, and each Latin upper case index gets a factor of the spin
connection according to
\begin{equation}
\qquad \nabla_\mu \, X^A_{\;\;B} = \partial_\mu \,X^A_{\;\;B}
     + \omega^{A}_{\;\;C \mu}\,X^C_{\;\;B}
     - \omega^{C}_{\;\;B \mu} \, X^A_{\;\,C} \,.
\end{equation}

Under local Lorentz transformations, $\Lambda^{A}_{\;\;B}\,(x) $, the spin connections
transforms as
\begin{equation}  \label{omega-Lorentz-transf}
\qquad \omega'^{A}_{\;\;B \mu}=\Lambda^{A}_{\;\;\,C}\,\, \omega^{C}_{\;\;D \mu}\,\,\Lambda_{B}^{\;\;\,D}
\,+\, \Lambda^{A}_{\;\;\,C}\,\, \partial_{\mu}\,\, \Lambda_{B}^{\;\;\,C}\,,
\end{equation}
where $\Lambda_B^{\;\;A}$ is the inverse of  $\Lambda^{A}_{\;\;B}\,$.
The spin connection in teleparallel gravity is meant to represent only inertial or frame effects.
This means that there exists a class of frames relative to which the spin connection
vanishes, $\widetilde{\omega}^{C}_{\;\;D \mu}=0$. From this fact along with equation~%
\eqref{omega-Lorentz-transf} one has that in a general class of frames the spin connection
takes the form~\cite{Krssak:2015rqa}
\begin{equation}  \label{omega-general}
\qquad \omega^{A}_{\;\;B \mu} = \Lambda^{A}_{\;\;\,C}\,\, \partial_{\mu}\,\,\Lambda_{B}^{\;\;\,C}\,.
\end{equation}

In  covariant $f(T)$ gravity theories, instead of the Levi-Civita connection,
one uses the Weitzenb\"{o}ck connection which is given by
\begin{equation} \label {connection}
\Gamma^{\rho}_{~\nu \mu}=h_{A}^{~\rho}\left(\partial_{\mu}{h^{A}_{~\nu}}
+\omega^{A}_{~B \mu} h^{B}_{~\nu}\right)\,,
\end{equation}
with inverse relation
\begin{equation}
\omega^{A}_{\;B \mu} = h^{A}_{\;\rho}\,\, \partial_{\mu}  h^{\rho}_{\;B} +
h^{A}_{\;\rho} \,\,\Gamma^{\rho}_{\;\,\nu \mu}\, h^{\nu}_{\;B} \equiv
h^{A}_{\;\rho} \, \nabla_{\mu} \, h_{B}{}^{\rho} \,,
\end{equation}
where $\nabla_{\mu}$ denotes the covariant derivative. An immediate consequence
of this definition for the covariant derivative
is that, for the spin connection~\eqref{omega-general} there is no curvature
but a nonzero torsion
\begin{equation}
T^{\rho}_{~\mu\nu}=h_{A}^{~\rho} T^{A}_{~\mu\nu}=\Gamma^{\rho}_{~\nu \mu}
-\Gamma^{\rho}_{~\mu \nu}\,,
\end{equation}
where in terms of the tetrad fields, the torsion tensor is defined
by~\cite{Aldrovandi-Pereira-book}
\begin{equation} \label{Def_Torsion}   
 T^{A}_{~\mu\nu}\equiv \partial_{\mu}h^{A}_{~\nu}
 -\partial_{\nu}h^{A}_{~\mu}+\omega^{A}_{~B\mu}\,h^{B}_{~\nu}
 -\omega^{A}_{~B\nu}\,h^{B}_{~\mu}\,.
\end{equation}

Now, if one further defines the so-called super-potential
\begin{equation} \label{5}
 S_{A}^{~\mu\nu}\equiv K^{\mu\nu}_{~~A}+h_{A}^{~\mu} \,T^{\theta\nu}_{~~\theta}-h_{A}^{~\nu}\,T^{\theta\mu}_{~~\theta}\,,
\end{equation}
where
\begin{equation}  \label{6}
 K^{\mu\nu}_{~~A}\equiv -\frac{1}{2}\left(T^{\mu\nu}_{~~A}
 -T^{\nu\mu}_{~~A}-T_{A}^{~\mu\nu}\right)
\end{equation}
is the contorsion tensor, we can define the torsion scalar
\begin{equation} \label{3}
 T \equiv \frac{1}{2} S_{A}^{~\mu\nu}\,T^{A}_{~\mu\nu}\,.
 \end{equation}
This scalar is used in the Lagragian density for the formulation of the covariant
$f(T)$ gravity theory, whose action is defined by
\begin{equation} \label{Action}  
S=\int{d^{\,4}x \,\,h\, \left[\frac{f(T)}{2 \kappa}+\mathcal{L}_{m}\right]}\,,
\end{equation}
where  $h=\det(h^{A}_{~\mu})$, $\kappa=8\pi G$, $f(T)$ is an arbitrary function
of torsion scalar $T$ and $\mathcal{L}_{m}$ is the Lagrangian density for the
matter field.%
\footnote{In the particular case $f(T)=T$ one has  the teleparallel equivalent
of general relativity (TEGR).} 

Varying the action~\eqref{Action} with respect to the vierbein and taking into
account equations~\eqref{Def_Torsion}~--~\eqref{3} one obtains the following
equations for the covariant $f(T)$ gravity:
\bea  \label{CovFieldE}
&& h^{-1}f_{T}\partial_{\nu}\left(h S_{A}^{~\mu\nu}\right)
+f_{TT} S_{A}^{~\mu\nu} \partial_{\nu}{T}-f_{T}h_{A}^{~\lambda}
T^{B}_{~\nu \lambda} S_{B}^{~\nu\mu} +\nonumber\\
&& f_{T}\omega^{B}_{~A \nu} S_{B}^{~\nu\mu}+\frac{1}{2}f(T) h_{A}^{~\mu}
=\kappa h_{A}^{~\nu} \Theta_{\nu}^{~\mu}\,,
\eea
where $f_T= df(T)/dT$, $f_{TT}= d^2f(T)/dT^2$, and $\Theta_{\nu}^{~\mu}$ is the
energy-momentum tensor of the matter fields defined as
\begin{equation}
\Theta_{\nu}^{~\mu}=\left(-\frac{1}{h}\frac{\delta{S_{m}}}{\delta h^{A}_{~\mu}}\right)
h^{A}_{~\nu},
\end{equation}
where $S_{m}=\int{d^{\,4}x \,h\,\mathcal{L}_{m}}$ is the matter action.

It should be emphasized that, unlike the usual formulation of $f(T)$ gravity,
where it is implicitly assumed that the spin connection $\omega^{A}_{~B\mu}$ vanishes,
in the derivation of the above field equations~\eqref{CovFieldE} a nonvanishing
$\omega^{A}_{~B\mu}$ is assumed from the outset. Actually, the presence the nonzero
spin connection term in field equations \eqref{CovFieldE} ensures the Lorentz covariance
of equations~\eqref{CovFieldE} (for details see Ref.~\cite{Krssak:2015oua}).
The price for this is that now we have to figure out a way to determine the nonvanishing
spin connection $\omega^{A}_{~B\mu}$, a point that we shall discuss in the following.

The covariant field equations~\eqref{CovFieldE} depend not only on the vierbein
but also on the nonvanishing spin connection $\omega^{A}_{~B\mu}$.%
\footnote{
In the usual $f(T)$ gravity  the field equations are differential equations for
the tetrads only.}
Thus, solutions to the covariant field equations can be found if one establishes a
procedure to suitably determine the spin connection $\omega^{A}_{~B\mu}$. In what
follows we shall briefly present a scheme figured out in Ref.~\cite{Krssak:2015oua}
to  determine $\omega^{A}_{~B\mu}$. It consists in starting from the arbitrarily
chosen initial tetrad field $h^{A}_{~\rho}$, define a \textit{reference frame},
$h^{A}_{(r)\mu}$, in which gravity is switched off, namely
\begin{equation} \label{hA-r-1}
h^{A}_{(r)\mu}\equiv h^{A}_{~\mu} \mid{}^{}_{\!\mbox{\small Grav}\,\longrightarrow \,0} \,.
\end{equation}
In other words, when the gravity is switched off the metric
\begin{equation} \label{hA-r-2}
\qquad g_{\mu\nu} = h^{A}_{(r)\mu}\,h^{B}_{(r)\nu}\,\eta_{AB}^{}
\end{equation}
reduces to the Minkowski metric.

Following the procedure of Ref.~\cite{Krssak:2015oua}, we then require the torsion
to vanish for this reference tetrad, i.e.
$T^{A}_{~~\mu \nu} (\, h^{A}_{(r)\mu}, \omega^{A}_{~B\mu})= 0 $. Thus,
to have the Lorentz covariant field equations we take the spin connection
for the field equations~\eqref{CovFieldE} as
\begin{equation} \label{omega-r}
\omega^{A}_{~B\mu}(h^{A}_{~\mu}) \equiv \omega^{A}_{~B\mu}(h^{A}_{(r)\mu})\,.
\end{equation}

The underlying basic idea behind this choice of spin connection is that the
torsion tensor \eqref{Def_Torsion} depends on both the tetrad and spin connection.
Thus, in general the torsion embodies the field
strength of gravity along with inertial effects. However, there exists a choice of
the spin connection for which the torsion captures only the strength of gravity.
This spin connection is obtained by using the reference tetrad $h^{A}_{(r)\mu}$
as defined by \eqref{hA-r-1} (or \eqref{hA-r-2}), and then by defining
$\omega^{A}_{~B\mu}$ according to \eqref{omega-r} (for more details see
Ref.~\cite{Krssak:2015oua}).

In practice, this can be made by taking, in the general expression for the
spin connection, (Eq.~(1.60) of Ref.~\cite{Aldrovandi-Pereira-book}),
\bea
 \omega^{A}_{~B \mu}=\frac{1}{2} h^{C}_{~\mu}\Big[ f_{B~C}^{~A}+T_{B~C}^{~A}
+f_{C~B}^{~A}+T_{C~B}^{~A} - f^{A}_{~B C}-T^{A}_{~B C}\Big] \,,
\eea
where $\,[h_A, h_B]= f^{\,C}_{\;\;AB}\, h_C\,$, and
the torsion $T^{A}_{~B C}=h_{B}^{~\mu}h_{C}^{~\nu}T^{A}_{~\mu\nu}=0$ for the
reference tetrad $h^{A}_{(r)\mu}$. In this way,  one has
\begin{equation} \label{spinc}
\omega^{A}_{~B \mu}=\frac{1}{2} h^{C}_{(r)\mu}\left[ f_{B~C}^{~A}(h_{(r)})
+f_{C~B}^{~A}(h_{(r)})-f^{A}_{~B C}(h_{(r)})\right],
\end{equation}
where the structure coefficients $f^{A}_{~B C}(h_{(r)})$ are calculated from
the expression (Eq.~(1.32) of Ref.~\cite{Aldrovandi-Pereira-book})
\be
f^{C}_{~A B}=h_{A}^{~\mu}h_{B}^{~\nu}\left(\partial_{\nu}{h^{C}_{~\mu}}
-\partial_{\mu}{h^{C}_{~\nu}}\right),
\label{CoeffA}
\ee
evaluated for the reference vierbein $h^{A}_{(r)\mu}\,$.

\section{G\"{o}del-type geometries}\label{Sec-GodelType}

In this section we present the main properties of the homogenous
G\"{o}del-type geometries, which we use in the next section.
We begin by recalling that G\"odel solution to the general relativity field
equations is a specific member of the large family of geometries,
whose general form in cylindrical coordinates, $(r, \phi, z)$,
is~\cite{Reb_Tiomno}
\begin{equation}  \label{G-type_metric}
ds^2 = [dt + H(r)d\varphi]^2 - D^2(r)d\varphi^2 - dr^2 - dz^2\,.
\end{equation}
The necessary and sufficient conditions for the G\"odel-type
geometries~(\ref{G-type_metric}) to be space-time homogeneous
(ST-homogeneous) are~\cite{Reb_Tiomno,RebAman}
\begin{equation} \label{GodelType1} 
\frac{H'}{D}  =  2\Omega \qquad \; \text{and} \qquad \;
\frac{D''}{D}  =  m^{2},
\end{equation}
where the prime denote derivative with respect $r$, and the parameters $(\Omega,m)$
are constants such that $-\infty\leq m^{2}\leq\infty$ and $\Omega^{2}>0$.
The ST-homogeneity is insured by the fact that G\"odel-type space-times admit
either a group $G_{5}$ or $G_{7}$ of isometries acting transitively on the
whole space-time~\cite{RebAman,TeiRebAman}.

Another important property of G\"{o}del-type geometries is the existence of an
irreducible set of isometrically nonequivalent classes of ST-homogeneous G\"odel-type,
which are given by~\cite{Reb_Tiomno} (see also Ref.~\cite{Fonseca-Neto:2013rna})
\begin{enumerate}
\item[\bf i.] Hyperbolic, in which $m^{2} = \mbox{const} > 0$ and
\begin{equation} \label{HGT-hyperb}
H =\frac{4\,\Omega}{m^{2}} \,\sinh^2 \left(\frac{mr}{2}\right), \;\;\;\;
               D=\frac{1}{m} \sinh\,\left(mr\right)\,;
\end{equation}
\item[\bf ii.] Linear, in which $m=0$ and
\begin{equation} \label{HGT-linear}
 H = \Omega r^{2},  \;\;\;\; D=r  \,,
\end{equation}
\item[\bf iii.] Trigonometric,  where $m^{2}= \mbox{const} \equiv - \mu^{2} < 0 $\
and
\begin{equation} \label{HGT-circul}
 H = \frac{4\,\Omega}{\mu^{2}} \,\sin^2 \left(\frac{\mu r}{2}\right),
                            \;\;\;\; D=\frac{1}{\mu} \sin\,\left(\mu r\right)\,.
\end{equation}
\end{enumerate}
Accordingly,  the ST-homogeneous G\"odel-type metrics are characterized by the
essential parameters $m^2$ and  $\Omega$.
Incidentally, G\"{o}del's solution of Einstein's equations is a
a particular member of hyperbolic class ($m^2 > 0$) in which $m^2=2\Omega^2$

To examine the causality features of the ST-homogeneous G\"odel-type metrics
we  rewrite the line element~(\ref{G-type_metric}) as
\begin{equation} \label{G-type_metric2}
ds^2=dt^2 +2\,H(r)\, dt\,d\phi -dr^2 -G(r)\,d\phi^2 -dz^2 \,,
\end{equation}
where $G(r)= D^2(r) - H^2(r)$.
In this form it is straightforward to show the existence of closed time-like
curves (CTC), i.e. make explicit the violation of causality in
ST-homogeneous G\"odel-type space-times. Indeed, from Eq.~(\ref{G-type_metric2})
one has that  the circles defined by $t, z, r = \text{const} $ with $r>0$
are closed timelike curves whenever $G(r) < 0$ (the line element $ds^2$ becomes
spacelike, and its integral curves are closed).
Thus, for the  hyperbolic class ($m^2 >0$) one finds that $G(r)= D^2(r) - H^2(r)$
is given by
\begin{equation}  
G(r) = \frac{4}{m^2} \, \sinh^2 \left(\frac{mr}{2}\right)
\left[\left( 1-\frac{4\Omega^2}{m^2}\right)\,
\sinh^2 \left(\frac{mr}{2}\right)+1 \right]\,.
\end{equation}
Therefore for $0 < m^2 < 4\Omega^2$ there is a finite critical radius $r_{c}$
defined by $G(r)=0$   
given by
\be  \label{r-critical}
\sinh^2\left(\frac{m r_{c}}{2}\right)=\left(\frac{4 \Omega^2}{m^2}-1\right)^{-1},
\ee
and such that $G(r)>0$ for $r<r_{c}$ and $G(r)<0$ for $r>r_{c}$.
Therefore, the circles $t$, $r$, $z = \mbox{const}$ in the circular band
with $r>r_c$ are CTC's.
One particularly important case in the range $0<m^2<4\Omega^2$ of the hyperbolic
class is the G\"{o}del metric ($m^2=2\Omega^2$), for the which there is a finite
critical radius and therefore breakdown of causality in the form of CTC's.
Another important G\"{o}del type metric in the hyperbolic family is
defined by $m^2=4\Omega^2$. Indeed, in this case from \eqref{r-critical}
one has that the critical radius goes to infinity, $r_{c} \rightarrow \infty$,
and therefore there is no violation of causality, since $G(r)> 0 $ for all
$0 < r < \infty$. An Einstein´s field equations solution of this specific
type was found in Ref.~\cite{Reb_Tiomno}.

For the linear family ($m=0$) one easily finds that $G(r)= D^2(r) - H^2(r)$
is given by
\be
G(r)=r^2 \left(1-\Omega r\right)\left(1+\Omega r\right).
\ee
Similarly to the hyperbolic class, for this family there is a critical
radius [$\,G(r)=0\,$] given  by $r_{c}=1/\Omega$ such that for any
radius $r > r_c$ the inequality $G(r)<0$ holds, and thus circles
defined by $t, z, r = \text{const}$ are CTC's.

Finally, for the trigonometric class $m^2 < 0$ one finds that $G(r)$
reduces to
\be  \label{Trig}
G(r)=\frac{4}{\mu^2}\sin^2\left(\frac{\mu r}{2}\right)\left[1
-\left(1+\frac{4\Omega^2}{\mu^2}\right)\sin^2\left(\frac{\mu r}{2}\right)\right],
\ee
but now differently from the other two class, there is an infinite sequence
of alternating causal [$\,G(r) > 0\,$] and noncausal [$\,G(r) < 0\,$] regions
(circular bands) in the section $t, z, r = \text{const} $ (with $r>0$) without
and with noncausal G\"{o}del's circles, depending on the value of $r$.
Thus, e.g., if $G(r) < 0$ for a certain range $ r_1 < r < r_2$ noncausal
circles exist, whereas  for $r$ in the next circular region $r_2 < r < r_3$
for which $G(r)>0$ no such noncausal G\"{o}del's circles
exist~\cite{Reb_Tiomno,Fonseca-Neto:2013rna}.

To close this section, we mention that throughout this work by non-causal and
causal solutions we mean solutions with and without violation of causality
of G\"odel-type, i.e., with and without the above-discussed G\"odel's noncausal
circles.

\section{G\"{o}del-type solutions in the covariant $f(T)$ gravity}
\label{GodelTypeinCovF(T)}

The aim of this section is twofold. First, following the scheme devised
in Ref.~\cite{Krssak:2015oua} we determine the spin connection associated with
a Lorenz tetrad of G\"{o}del-type space-time geometries~\eqref{G-type_metric},
which allow for the field equations for these space-times in the covariant
$f(T)$ gravity.
Second,  we take a combination of a perfect fluid with electromagnetic plus
a scalar fields as source, and determine a general G\"odel-type solution
of the covariant $f(T)$ field equations, and discuss its main important
properties.

\subsection{Field equations}

At an arbitrary point G\"odel-type space-time manifold we choose the
following tetrad basis
$\theta^A = h^{A}_{~\mu}\,dx^\mu\,$:
\begin{eqnarray}
\theta^0 &=& dt + H(r)\,d\varphi\,, \quad  \theta^1 = dr\,, \label{one_forms1} \\
\theta^2 &=& D(r)\,d\varphi\,,        \,\quad \qquad \theta^3 = dz \label{one_forms2} \,,
\end{eqnarray}
relative to which the G\"odel-type line element~\eqref{G-type_metric} clearly
takes the form
\begin{equation}  \label{G-type_metric3}
ds^2 = \eta^{}_{AB}\,\theta^A\theta^B =
\eta^{}_{AB}\, h^{A}_{~\mu}\,h^{B}_{~\nu}\, dx^\mu\, dx^\nu\,,
\end{equation}
$\eta_{AB} = \mbox{diag} (+1, -1, -1, -1)$, and the components of the tetrad
fields $h^{A}_{~\mu}$ and $h_{A}^{~\mu}$ are given by

\begin{equation}  \label{Godeltetrad}
h^{A}_{~\mu}= \left( \begin{array}{cccc}
1 & 0 & H & 0  \\
0 & 1 & 0 & 0 \\
0 & 0 & D & 0 \\
0 & 0 & 0 & 1 \end{array} \right)
\quad \mbox{and} \quad
h_{A}^{~\mu}= \left( \begin{array}{cccc}
1 & 0 & -\frac{H}{D} & 0  \\
0 & 1 & 0 & 0 \\
0 & 0 & \frac{1}{D} & 0 \\
0 & 0 & 0 & 1 \end{array} \right).
\end{equation}

{}From equations~\eqref{G-type_metric3} and \eqref{Godeltetrad} one has that gravity
is switched off when  $H(r)\rightarrow 0$ and $D(r)\rightarrow r$, inasmuch as in this limit
the tetrads reduce to the tetrads of the Minkowski space-time in cylindrical coordinates,
namely $h^{A}_{\mbox{\tiny (M)}\mu}= \mbox{diag} (1,1,r,1)$.
Thus, following the scheme devised in Ref.~\cite{Krssak:2015oua} the \textit{reference frame},
$h^{A}_{(r)\mu}$, defined by
\begin{equation}
h^{A}_{(r)\mu} \equiv h^{A}_{~\mu} \mid{}^{}_{\!\mbox{\small Grav}\, \longrightarrow \,0} \,,
\end{equation}
is given by
\begin{equation} \label{RTetrad}
 h^{A}_{(r)\mu} = h^{A}_{\mbox{\tiny (M)}\mu} = \mbox{diag} (1,1,r,1) \,.
\end{equation}
For this reference frame by using Eq.~\eqref{CoeffA} we obtain
the following nonvanishing components of the coefficients of anholonomy:
\begin{equation}
f^{\hat{2}}_{~\hat{2}\hat{1}}=-f^{\hat{2}}_{~\hat{1} \hat{2}}=\frac{1}{r},
\end{equation}
and then, by using Eq.~\eqref{spinc}, one finds that the  nonzero spin
connection  associated to the reference frame \eqref{RTetrad} is given
by
\begin{equation}
\omega^{\hat{2}}_{~\hat{1} 2}=-\omega^{\hat{1}}_{~\hat{2} 2}=1,
\label{cylinspinc}
\end{equation}
where here and in what follows the hat over the numerical indices is used to
denote that the corresponding digits are tetrad indices.

To have the Lorentz covariant field equations we take the spin
$\omega^{A}_{~B\mu}(h^{A}_{~\mu})=\omega^{A}_{~B\mu}(h^{A}_{(r)\mu})$.
Thus, from equations~\eqref{Godeltetrad} and~\eqref{cylinspinc}
the field equations~\eqref{CovFieldE} reduce, for G\"odel-type spacetimes,
to %
\footnote{We note that the additive term $ D^{-1} f_{TT}T'$ in equations \eqref{FQ1}
and~ \eqref{FQ6} do not arise in the usual formulation of $f(T)$ gravity. For comparison
we refer the readers to Ref.~\cite{Liu:2012kka}, where similar field equations were derived
in the context of the usual $f(T)$ gravity.}

\bea
&& \left[\frac{HT'}{2 H'}-\frac{D''}{D}+T\right]f_{T}+\left[\frac{HT}{H'}
-\frac{D'}{D}\right]f_{TT}T'+\frac{1}{D} f_{TT}T' + \nonumber\\
&&\frac{f}{2}
=\kappa\,\Theta_{0}^{~0}\,,\label{FQ1} \\ 
&& \left[\frac{D^2 T'}{2 H'}+H T\right] f_{T}+\frac{H' T'}{2} f_{TT}-\frac{H}{2}f=\kappa\,\left(\Theta_{2}^{~0}-H\Theta_{0}^{~0}\right)\,,\label{FQ2}\nonumber \\\\
&& -T f_{T}+\frac{f}{2} = \kappa\,\Theta_{1}^{~1},\label{FQ3}\\
&&-\frac{T'}{ H'}\left[\frac{f_{T}}{2}+Tf_{TT}\right]=\kappa\,\Theta_{0}^{~2}\,,
\label{FQ4} \\
&& -T f_{T}+\frac{f}{2}=\kappa\,\left(\Theta_{2}^{~2}-H\Theta_{0}^{~2}\right)\,,\label{FQ5}\\
&& \frac{1}{D} f_{TT}T'-\frac{D'}{D} T' f_{TT} -\frac{D''}{D}f_{T}+\frac{f}{2}
=\kappa\,\Theta_{3}^{~3}\,,
\label{FQ6}
\eea
where we have used that, from Eq.~\eqref{3} along with Eqs.~\eqref{5}, \eqref{6}
and \eqref{Def_Torsion}, the torsion scalar is given by
\begin{equation}  \label{ScalarTorsion}
T=\frac{1}{2}\left(\frac{H'}{D}\right)^2\,.
\end{equation}

Since in this paper we are interested in ST-homoneneous G\"odel-type space-times,
the homogeneity conditions~\eqref{GodelType1} holds, and therefore torsion
scalar is  given by 
\begin{equation}  \label{STHomogenous}
T = 2 \Omega^2\,,
\end{equation}
and the field equations~\eqref{FQ1} to~\eqref{FQ6} reduce to
\bea
&& \left(2\Omega^2 - m^2 \right)f_{T} +\frac{f}{2} =\kappa\,\Theta_{0}^{~0}\,,\label{FQ1A} \\
&& 2\Omega^2 H f_{T} -\frac{H}{2}\,f=\kappa\left(\Theta_{2}^{~0}-H\Theta_{0}^{~0}\right),\label{FQ2A}
\eea \vspace{-5mm} \bea
&&  -2\Omega^2 f_{T}+\frac{f}{2}=\, \kappa\,\Theta_{1}^{~1}=\,\kappa\,\Theta_{2}^{2} \,,\label{FQ3-5A}\\
&& \qquad \Theta_{0}^{~2} = 0 \,,\label{FQ4A}\\
&& -m^2 f_{T} + \frac{f}{2}=\kappa\,\Theta_{3}^{~3} \label{FQ6A}\,.
\eea
Since the homogeneous G\"odel-type metric~\eqref{G-type_metric} and tetrads
$h^{A}_{~\mu}$ [Eq.\eqref{Godeltetrad}] are determined by the parameters $m^2$ and
$\Omega$, solving equations~\eqref{FQ1A}~--~\eqref{FQ6A} for these parameters
one finds
\bea \label{SFQ3}
&& m^2=\frac{\kappa}{f_{T}}\left(\Theta_{0}^{~0}+\Theta_{1}^{~1}-2
\Theta_{3}^{~3}\right),\label{SFQ2}\\
&&\Omega^2=\frac{\kappa}{2 f_{T}}\left(\Theta_{0}^{~0}-\Theta_{3}^{~3}\right),
\label{SFQ1}
\eea
subject to the constraints
\begin{eqnarray} \label{Contraint-eq1}
&&f  = 2\kappa\left(\Theta_{0}^{~0}+\Theta_{1}^{~1}-\Theta_{3}^{~3}\right)\,,\\
&&\Theta_2^{~0} - \,H (\Theta_0^{~0} - \Theta_1^{~1})=\Theta_{0}^{~2} = 0 \,,
\label{Contraint-eq2}
\end{eqnarray}
where the functions $f$ and $f_{T}$ are  evaluated at $T=2\Omega^2$, and where we assume
that $f_{T}>0$ to ensure that the effective Newton constant does not change its sign.

It should be emphasized that in the field equations~\eqref{SFQ3}~--~\eqref{Contraint-eq2}
no specific energy momentum tensor have been used. In the next subsection, a concrete matter
content will be considered in the search for ST-homogeneous G\"{o}del-type solutions
of the covariant $f(T)$  field equations. To this end, we shall take a combination of a perfect
fluid with scalar field along with an electromagnetic field. Clearly, the scalar and electromagnetic
field sources have to fulfill, respectively, the Klein-Gordon and Maxwell equations in the
curved G\"odel-type background geometry.

\subsection{Solutions}
In this Section we discuss  ST-homogeneous G\"odel-type solutions of the covariant
$f(T)$ gravity for specific matter sources. To this end, we take combination of scalar
and electromagnetic fields with a perfect fluid as a matter source, find a general
solution, and examine several special G\"odel-type  solutions.
Thus, in the local frame defined by equations \eqref{one_forms1}~and~\eqref{one_forms2}
the energy-momentum tensor $\Theta_{AB}$ for the combined fields has the form
\begin{equation} \label{TotalEMT}
\Theta_{A B}=\overset{\mathbf{pf}}{\Theta}_{A B}+\overset{\mathbf{sf}}{\Theta}_{A B}+\overset{\mathbf{ef}}{\Theta}_{A B},
\end{equation}
where $\overset{\mathbf{pf}}{\Theta}_{A B}$, $\overset{\mathbf{sf}}{\Theta}_{A B}$ and $\overset{\mathbf{ef}}{\Theta}_{A B}$ are, respectively, the energy-momentum tensors
of a perfect fluid, a scalar field and an electromagnetic field, which are given by
\bea
&& \overset{\mathbf{pf}}{\Theta}_{A B}=\left(\rho+p\right) u_{A} u_{B}-p\,\eta_{A B}\,,\label{PF} \\
&& \overset{\mathbf{sf}}{\Theta}_{A B}=\nabla_{A}\phi \,\nabla_{B}\phi-\eta_{A B}\left[\frac{1}{2}
\eta^{C D} \nabla_{C}\,\phi\, \nabla_{D}\, \phi\right]\,, \label{SF} \\
&& \overset{\mathbf{ef}}{\Theta}_{A B}=-F_{A}^{~C} F_{B C}+\frac{1}{4} \eta_{A B}\,
F_{C D} F^{C D}\,,\label{EF}
\eea
where $u_{A} = \delta_A^{0}$ is the four-velocity,  $\rho$ and $p$ are the energy density
and pressure of the perfect fluid, subject to the weak energy condition (WEC) $\rho>0$ and
$\rho+p>0$, and where $\nabla_{A}\phi$ denotes the covariant derivatives relative to the
local basis $\theta^A = h^{A}_{~\alpha} \, dx^\alpha$.

The massless scalar field $\phi$ fulfills the Klein-Gordon equation
\be
\Box{\phi}=\eta^{AB}\nabla_{A}\nabla_{B}\,{\phi}=0\,,
\ee
a solution of which can be written as $\phi=\phi(z)=s\left(z-z_{0}\right)$, where
$s, z_0 = \mbox{const}$~\cite{Reb_Tiomno}. Thus,  the non-vanishing
components of the energy-moment tensor  are
\begin{equation}
\overset{\mathbf{sf}}{\Theta}_{\hat{0} \hat{0}}=
-\overset{\mathbf{sf}}{\Theta}_{\hat{1} \hat{1}}=
-\overset{\mathbf{sf}}{\Theta}_{\hat{2} \hat{2}}=
\overset{\mathbf{sf}}{\Theta}_{\hat{3} \hat{3}}= \frac{s^2}{2}\,,\label{SF2}
\end{equation}

The electromagnetic field $F_{A B}$ satisfies the source-free Maxwell equations
\begin{equation} \label{MaxEqs}
\nabla_{A}{F^{A B}} = 0
\qquad \mbox{and} \qquad
\nabla_{A}{\left(\star F^{A B}\right)}=0\,,
\end{equation}
where $\star F^{A B}=\frac{1}{2}\varepsilon^{ABCD}F_{CD}$.
Following Ref.~\cite{Reb_Tiomno}, it can be shown that the electromagnetic field
tensor $F_{AB}$ given by
\begin{eqnarray} 
F_{\hat{0}\hat{3}} = - F_{\hat{3}\hat{0}} &=& e\,\sin[2\Omega(z-z_0)]\,, \\
\: F_{\hat{1}\hat{2}} = - F_{\hat{2}\hat{1}} &=& e \,\cos[2\Omega(z-z_0)]\,,
\end{eqnarray}  
where $e$ is constant, satisfies Maxwell equations~\eqref{MaxEqs}.
Hence, the non-vanishing components of the associated energy-momentum tensor
$\Theta_{AB} = F_A^{\ C} \, F^{}_{BC} + \frac{1}{4} \eta^{}_{AB} F^{CD} F_{CD}$
are given by
\begin{equation} \label{EF2}
\overset{\mathbf{ef}}{\Theta}_{\hat{0} \hat{0}}=
\overset{\mathbf{ef}}{\Theta}_{\hat{1} \hat{1}}=
\overset{\mathbf{ef}}{\Theta}_{\hat{2} \hat{2}}=
-\overset{\mathbf{ef}}{\Theta}_{\hat{3} \hat{3}}=\frac{e^2}{2}.
\end{equation}

Thus, taking into account equations~\eqref{PF}, \eqref{SF2} and \eqref{EF2},
one has that for the combined-fields matter source~\eqref{TotalEMT}, and
therefore from equation~\eqref{SFQ2}, \eqref{SFQ1} and \eqref{Contraint-eq1} one
has that a general solution to the field equations can be written as%
\footnote{Clearly we have used that the coordinate components of the energy momentum
tensor are given by $\Theta_{\mu}^{~\nu}=  h_{B}^{~\nu}\, h_{\mu}^{~A} \,\Theta_A^{~B}$.}
\bea
&& m^{2}=\frac{\kappa}{f_{T}}\left(\rho+p+2s^2-e^2\right)\,,\label{m2} \\
&& \Omega^2=\frac{\kappa}{2 f_{T}}\left(\rho+p+s^2\right)\,, \label{Omega2}\\
&& f=\kappa\left( 2\rho+3s^2-e^2\right)\,, \label{fC}
\eea
and equations~\eqref{Contraint-eq2} are identically satisfied.

Since ST-homogeneous G\"odel-type geometries~\eqref{G-type_metric} and
tetrads~\eqref{Godeltetrad} are characterized by the two essential parameters
$m^2$ and  $\Omega$, the above equations~\eqref{m2} and \eqref{Omega2} along
with the constraint equation~\eqref{fC} make explicit how the covariant $f(T)$
gravity  specifies a pair of parameters $(m^2, \Omega^2)$, and therefore determines
a general ST-homogeneous G\"odel-type solution for the combined-fields matter
source~\eqref{TotalEMT}.
In the remainder of this Section we discuss the most significant special cases of this
general solution and some important characteristics of these G\"odel-type solutions in
the covariant $f(T)$  gravity.

A first important point regarding the above combined field general solution is that
from equation~\eqref{m2} we clearly have solutions in the hyperbolic~%
\eqref{HGT-hyperb}, linear~\eqref{HGT-linear} or trigonometric~\eqref{HGT-circul}
families of G\"odel-type space-times.

Since $f_T>0$ solutions are in the hyperbolic class when $\rho+p+2s^2-e^2>0$. We note,
however, that from equations~\eqref{m2} and~\eqref{Omega2} there is an upper bound
for $m^2$ since
\begin{equation}
m^2-4\Omega^2=-\frac{\kappa}{f_{T}}\left(\rho+p+e^2\right) \leq 0.
\end{equation}
{}From equations~\eqref{m2} and~\eqref{Omega2} it is easy to show that the limiting
bound solution $m^2=4 \Omega^2$ takes place for single scalar field of amplitude $s$
as the matter source. Besides, according to Eq.~\eqref{r-critical} of
Section~\ref{Sec-GodelType} the critical radius goes to infinity, $r_{c} \rightarrow \infty$,
and therefore there is no violation of causality for this particular solution.
Apart from this particular case, from equation~\eqref{r-critical} along with
\eqref{m2} and \eqref{Omega2} one has that for all solutions in the hyperbolic class
with $m^2<4 \Omega^2$ there is a finite critical radius, beyond which causality is
violated, given by
\begin{equation} \label{GodelR2}  
r_{c}=\frac{2\, ({f_T})^{\frac{1}{2}}\, \sinh^{-1} \left(\sqrt{
\frac{\rho+p+2s^2-e^2}{\rho+p+e^2}}\right)}
{\left[ \kappa\left(\rho+p+2 s^2-e^2\right)  \right]^{1/2} } \,,
\end{equation}
As one would expect from the outset this critical radius depends upon the
$f(T)$ gravity theory and the matter content.

A particularly important case in the hyperbolic family is the G\"{o}del
spacetime, in which  $m^2=2 \Omega^2$. From equations~\eqref{m2} and~\eqref{Omega2}
there are two possible matter sources that give rise to G\"odel solution
in covariant $f(T)$ gravity. First, a perfect fluid of density
$\rho$ and pressure $p$ subject to the WEC and with $f_T>0$. In this case $s^2=e^2=0$
and therefore $m^2=2 \Omega^2$ holds.%
The fact that all perfect fluid solutions subject to the WEC  are isometric to G\"odel
geometry extends to the context of covariant $f(T)$ gravity with $f_T>0$ the Bampi
and Zordan theorem~\cite{BampiZordan78} demonstrated in the framework of GR. 
Second, for a simple combination of the above scalar and electromagnetic fields
($\rho=p=0$) from equations~\eqref{m2} and~\eqref{Omega2} one has
\begin{equation}
m^2-2\Omega^2=\frac{\kappa}{f_{T}}\left(s^2-e^2\right)\,,
\end{equation}
Therefore, for  $s^2=e^2\neq 0$ G\"{o}del geometry is again recovered.

Regarding the linear and trigonometric families of G\"odel-type space-times, from Eq.\eqref{m2}
we clearly have that to have either one of these classes the existence of the electromagnetic
field component is required.
The solutions in the linear class ($m^2=0$) are obtained when $\rho+p+2s^2-e^2=0$.
Thus, for example, a simple combination of the the above scalar and electromagnetic fields
such that $e^2=2 s^2$ gives rise to a  solution in the linear class.
Since $\rho+p>0$ from WEC,  in general solutions in the linear class for the combined fields
source arise only when $e^2 \geq 2 s^2$.
According to Section~\ref{Sec-GodelType} and similarly to the hyperbolic class,
for solutions belonging to the linear family there is a critical radius $r_{c}=1/\Omega$
such that for any  $r > r_c$ the G\"odel circles (CTC) defined by $t, z, r = \text{const}$
arise.

Finally, solutions in the trigonometric class ($m^2=-\mu^2<0$) arise when
$\rho+p+2s^2-e^2<0$. Very simple examples of solutions in this class come
about when $\rho=p=0$ and $e^2 > 2 s^2$, but clearly we can have solutions with all
the above combined field matter components.
As for the violation of causality, according to Section~\ref{Sec-GodelType}
for the trigonometric class there is an infinite sequence of alternating causal and
noncausal circular bands in the section $t, z, r = \text{const}$, $r>0$,
with and without noncausal G\"{o}del's circles, depending on whether $\,G(r) < 0\,$
or $\,G(r) > 0\,$. 

\section{Concluding remarks} \label{Conclu}

Despite the undeniable success of the general relativity theory, there has
been a great deal of recent works on the so-called modified gravity
theories. In the cosmological modeling  this is motivated by
the fact that these modified theories furnish an alternative way to
account for the recent expansion of the Universe with no need to
invoke the dark energy matter component.   
Several features of a family of these modified gravity theories,
known as $f(T)$ gravity, have been discussed recently in a number
of articles.
Despite this noticeable interest in this new gravitational theory, it appears
to have been overlooked  in the recent literature, that the $f(T)$ field
equations are not invariant under local Lorentz transformations, unless
the $f(T)$ is a linear function of $T$, which is the teleparallel equivalent
of general relativity (TEGR).
This means that apart from the TEGR, different choices of the Lorentz
tetrads give rise to different field equations.
Relaxing a particular choice of the spin connection often made implicitly in
the literature a locally Lorentz covariant $f(T)$ gravity theory has been
devised recently~\cite{Krssak:2015oua}.

A well-behaved chronology and causality are so essential in the special
relativity theory that they are simply incorporated into the theory from
its bare formulation.
General relativity (GR) inherits locally the chronology and causality
features of special relativity.
On nonlocal scale, however, important differences emerge and Einstein's
field equations admit solutions to its field equations with closed timelike
world lines, despite its local invariance under Lorentz transformations
that ensures locally the validity of the causality.

Even though different choices of local Lorentzian tetrads leave the metric
invariant, suggesting at first sight a well-defined chronology and
well-behaved causal structure, in the  standard formulation of $f(T)$ gravity
every different Lorentz tetrads  give rise to different field equations,
and therefore represents, in general, a different theory.
Hence, for a given (fixed) Lorentz frame, which determines fixed set of
field equations (a fixed theory), there is no Lorentz transformation freedom.
This means that unlike general relativity, a $f(T)$ gravity does not inherit
locally the chronology and the well-behaved causal structure of special
relativity.

This local causality problem seems to have been overcome in the Lorentz
covariant $f(T)$ gravity~\cite{Krssak:2015oua}, since in the
new formulation of $f(T)$ theory the Lorentz transformations do not
change neither the metric nor the field equations.
The nonlocal question, however, is left open and violation of causality
may arise in the context of the covariant $f(T)$ gravity theory.

Since homogeneous G\"odel-type geometries necessarily lead to the
existence of closed timelike circles --- an explicit manifestation of
violation of causality on nonlocal scale ---, a plausible 
way to answer this question is by investigating whether the covariant
$f(T)$ gravity theories permit G\"odel-type solutions.
Furthermore, if gravity is to be governed by a covariant $f(T)$ theory
there are a number of issues ought to be examined in its context,
including the question as to whether these theories permits G\"odel-type
solutions, or would remedy this causality problem by ruling them out.

In this paper, to look into the potentialities, difficulties and limitations
of the covariant $f(T)$ gravity theories, we have undertaken one of these questions
and examined whether these theories admit homogeneous G\"odel-type solutions
for a number of matter sources.
We have taken a combination of a perfect fluid with electromagnetic plus
a scalar fields as a matter source, and determined a general G\"odel-type
solution.
This general solution contains special solutions whose essential parameter,
$m^2$, can take the sign that defines any class of homogeneous G\"odel-type
geometries, namely hyperbolic family ($m^2>0$), trigonometric class ($m^2 < 0$)
and linear class ($m=0$).
We have found that solutions in the trigonometric and linear classes are
permitted only for the combined matter sources that includes an electromagnetic
field matter component.
We have extended to the context of covariant $f(T)$  gravity the
so-called Bampi-Zordan theorem, which ensures that any perfect-fluid
ST-homogeneous G\"odel-type solution in this theory $f(T)$ gravity is
isometric to G\"odel metric, and therefore defines the same set of
G\"odel tetrads $h_A^{~\mu}$ up to a Lorentz transformation.
We have also shown  that the single massless scalar field generates
the only ST-homogeneous G\"odel-type solution with no closed timelike
curves.

To conclude, we underline that it emerges from our results that even
though the covariant $f(T)$ gravity restores local Lorentz covariance
of the field equations and the validity of the causality principle locally,
the bare existence of the G\"odel-type solutions makes apparent that
this covariant formulation of $f(T)$ gravity does not avoid non-local
violation of causality in the form of closed timelike curves that are
permitted in general relativity.

\begin{acknowledgements}
M.J. Rebou\c{c}as acknowledges the support of FAPERJ under a CNE E-26/102.328/2017
grant and also thanks CNPq for the grant under which this work was carried out.
G. Otalora acknowldeges DI-VRIEA for financial support through Proyecto Postdoctorado
$2017$ VRIEA-PUCV. We  are very grateful to A.F.F. Teixeira for indicating important omissions
and misprints.
\end{acknowledgements}


\end{document}